# The Level I Multiverse is not the same as the Level III Multiverse

Alan McKenzie

*Lately of University Hospitals Bristol NHS Foundation Trust, Bristol UK*[*]

**Abstract**

Anthony Aguirre and Max Tegmark have famously speculated that "the Level I Multiverse is the same as the Level III Multiverse". By this, they mean that the parallel universes of the Level III Multiverse can be regarded as similar or identical copies of our own Hubble volume distributed throughout the whole of our (possibly infinite) bubble universe. However, we show that our bubble universe is in a single quantum eigenstate that extends to regions of space that are receding from each other at superluminal velocities because of general relativistic expansion. Such a bubble universe cannot accommodate Hubble volumes in the different orthogonal eigenstates required by the Level III Multiverse. Instead, quantum uncertainty arises from large numbers of alternative bubble universes in Hilbert space, isolated from each other. The conclusion of the paper is that the Level I Multiverse is not the same as the Level III Multiverse.

## 1. "The Level I Multiverse is the same as the Level III Multiverse"

In the universes in which I am writing this, Anthony Aguirre is the first author of a paper that he wrote with Max Tegmark [1] where they decided upon the author order from the outcome of a single (unspecified) quantum measurement. They point out that this author order is found in "exactly half of all otherwise-indistinguishable worlds spread throughout space"; in the other half, the order is reversed. The controversial part of their claim, of course, is not the existence of parallel universes where the author order for their paper is not alphabetical, but their suggestion that these parallel universes might all inhabit regions of the same Euclidean three-dimensional space (three-space).

The idea of parallel universes emerged, of course, from Hugh Everett's doctoral thesis, in which he showed that Niels Bohr's concept of wave function collapse is unnecessary [2]. If you take the wave function of the whole universe and simply follow its unitary evolution according to the Schrödinger equation, it rapidly grows into a grand superposition of orthogonal states, each of which we can call a parallel universe. So the default location of parallel universes was not in three-dimensional Euclidian space but in infinite-dimensional Hilbert space.

---

[*] www.godel-universe.com

As soon as Bryce DeWitt announced Everett's work to the wider physics community in his 1970 review in *Physics Today*, [3], many critics disparaged the notion of such a plethora of universes, each featuring identical copies of ourselves ("schizophrenia with a vengeance" DeWitt admitted). Looking back, perhaps the deep reason why many found it difficult to accept the Many Worlds Interpretation (MWI) is that, in doing so, they would be signing up to the notion that our universe, including all of its self-aware inhabitants, is simply part of a vast mathematical structure that includes all of the other universes, manifest as mere sets of vectors in Hilbert space.

It was Tegmark who, in 1997, proposed that the universe is just such a mathematical structure [4], following it up a decade later with his Mathematical Universe Hypothesis [5]. This formalization of the notion that the universe *is* mathematics may indeed have contributed to the growing acceptance of MWI among physicists [6], [7].

However, for some physicists, the branching of MWI remained a major sticking point: they perceived MWI to be silent on (1) the question of what happens precisely at the point where the wave function branches (decoherence notwithstanding); and (2) how the "thickness" of a branch (the absolute square of its probability amplitude) translates into its universe "having a greater or lesser reality".

Replacing the "MWI tree" with parallel block universes – filaments that extend from the trunk to the furthest twigs and which populate branches in numbers proportional to the branch thicknesses – makes these two objections to MWI branching irrelevant [8]. (The motivation for introducing the parallel block-universe hypothesis, though, was different. An experiment can show that any *future* quantum event in the universe is nevertheless already in the *past* of an observer travelling at an appropriate velocity [9]. So, since the past cannot be changed, neither can the future – so there is only one future. This rules out branching as suitable topology for a universe, leaving the parallel block-universe hypothesis as a natural choice.)

The question of whether parallel worlds can exist within the three-dimensional space of our own universe was raised before Guth published his first paper on inflation [10]. As Ellis and Brundrit pointed out, "there is no need to postulate some hypothetical statistical ensemble – it exists in the infinite universe!" [11].

At this point, we need to be clear about what we mean by a "universe". Figure 1 highlights terms that will be important for us. The background grey represents the false vacuum which drives inflation [12], called the "inflaton field" in the diagram. Distance is in the horizontal direction and time is in the vertical direction. Two bubble universes have thermalized from the false vacuum [13] and are shown in white. Although these are expanding exponentially, their boundaries are asymptotically vertical because comoving coordinates have been used. The thermalized region is most commonly known as a *bubble universe*, although Guth prefers the term *pocket universe* in order to avoid the impression that the boundary is smooth like a bubble. Tegmark [5] has called our bubble universe a *Level I Multiverse* because he reserves the term "universe" for our Hubble volume[†]. So, in

---

[†] A Hubble volume is the spherical region, centred on an observer, beyond which the expansion of space away from the observer is greater than the speed of light. Since this is smaller than the observable universe (because, during the time light takes to reach us from the Hubble boundary, the

this picture, there are many (and, many argue, an infinite number of) such Hubble volumes – Tegmark universes – within our single bubble universe. It is this collection of Tegmark universes that he terms the "Level I Multiverse".

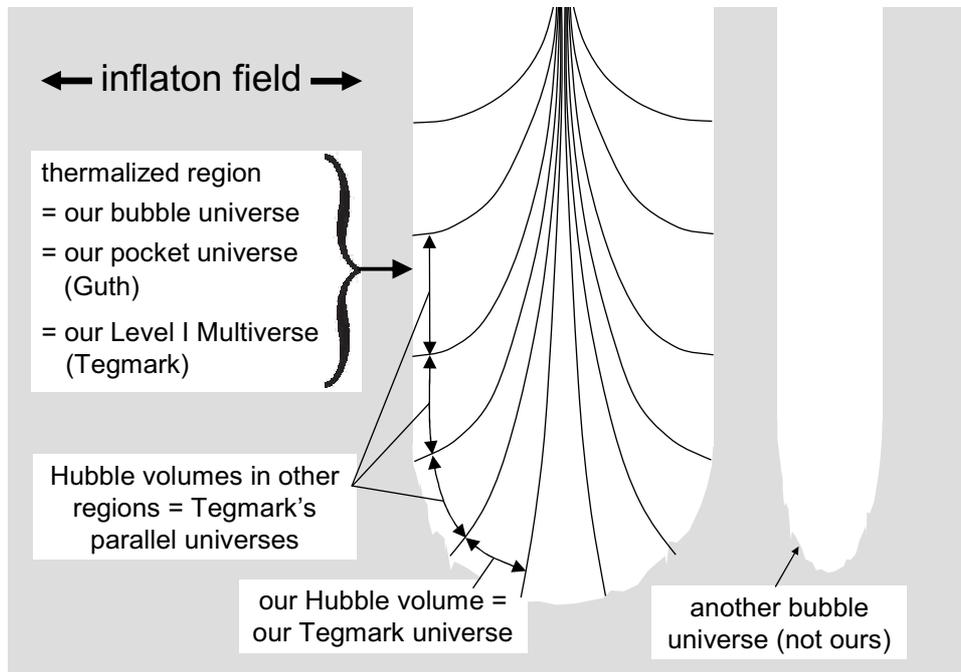

**Figure 1:** This is intended to clarify the usage of the word "universe" and other terms. The background grey represents the inflaton field and the two white "U" shapes are bubble universes. Comoving coordinates are used in the horizontal direction and time is in the vertical direction.

Garriga and Vilenkin [14] further developed the idea of there being parallel universes within our own three-space, and, in the context of inflationary cosmology, considered our own thermalized region, or bubble universe. The future light cone emanating from the spacetime point of origin of a bubble universe becomes its effective boundary, and space, as viewed from within the bubble universe, extends along the light cone asymptotically. Since the process of inflation, once it has started, continues indefinitely in generic models [15], [16], it is common to think of the bubble universe as extending spatially to infinity from the perspective of its inhabitants.

Of course, while some physicists may consider our own particular bubble universe to be spatially infinite, the volume that we can observe from within our universe is not, and, indeed, from arguments of the Bekenstein-bound type [17], our Hubble volume can be in one of only a finite number of different states. Garriga and Vilenkin [14] point out that there must be an infinite number of such finite volumes in our infinite universe: since the number of states available to our own Hubble volume is finite, there must be an infinite number of Hubble volumes throughout our universe that are in identical states to that of our own. They liken this ensemble of finite volumes to

---

space from where the light was emitted has expanded beyond it), a case may be made for using the observable universe rather than the Hubble volume. However, it ultimately makes no difference to the discussion, and so we adopt Tegmark's usage.

the ensemble of universes in MWI, with the important difference that all of the finite volumes are, in their words, "unquestionably real".

In the paper cited at the beginning of this article, Aguirre and Tegmark [1] discuss just such an ensemble of generally non-contiguous finite volumes spatially distributed throughout our infinite bubble universe. They show that, if this ensemble of volumes, or regions, is populated with quantum outcomes in proportion to the absolute squares of their probability amplitudes, then observers in any one of these different regions will conclude that such outcomes occur in accordance with the Born rule.

If this model is correct, then it accounts for quantum uncertainty, as the authors claim. Indeed, they suggest that "the Level I Multiverse is the same as the Level III Multiverse" (where the universes of the Level III Multiverse are analogous to the worlds/universes of MWI). (As Aguirre and Tegmark point out, the very process of creating thermalized regions like our own bubble universe may well itself be a quantum one, but that, and related questions, should be addressed in a separate forum.) To illustrate the argument, imagine a large group of many identical regions in the bubble universe, each described by a state $|\psi\rangle$, in which a Stern-Gerlach experiment is about to determine the spin of a particle. Because of the particle's initial spin and the relative orientation of the detecting magnetic field, let us suppose that there is a ¾ chance of the particle being detected with an up-spin, $|\uparrow\rangle$, and a ¼ chance of it being detected with a down-spin, $|\downarrow\rangle$. After the measurement, the states of three quarters of the regions may each be written $|\psi_\uparrow\rangle \otimes |\uparrow\rangle$, and the remaining quarter may be written $|\psi_\downarrow\rangle \otimes |\downarrow\rangle$ where $|\psi_\uparrow\rangle$ is the state of a region after an up-spin has been observed and $|\psi_\downarrow\rangle$ is the state after a down-spin has been observed. Figure 2 shows four such regions after the experiment. In the Aguirre-Tegmark model, these regions are replicated in this proportion very many times across the bubble universe (actually, according to the authors, an infinite number of times, but, in order to avoid controversy over the meaning of ratios of infinities, we imagine a vast, but finite, bubble universe).

## 2. The regions of Aguirre and Tegmark are eigenstates

The question is: does this model hold true? Is it indeed possible to have ensembles of volumes distributed throughout the bubble universe, differing only in quantum outcomes where these outcomes are populated according to the absolute squares of their probability amplitudes as required by the Born rule?

An essential feature of the Aguirre-Tegmark model is that the bubble universe is a coherent mathematical structure. The word *coherent* is strictly unnecessary: it just means that the complete structure of the bubble universe is interlinked mathematically. The explicit attribute of coherence is unnecessary because that is implied by the notion of a structure anyway. Nevertheless, it is a useful reminder in the light of what comes later.

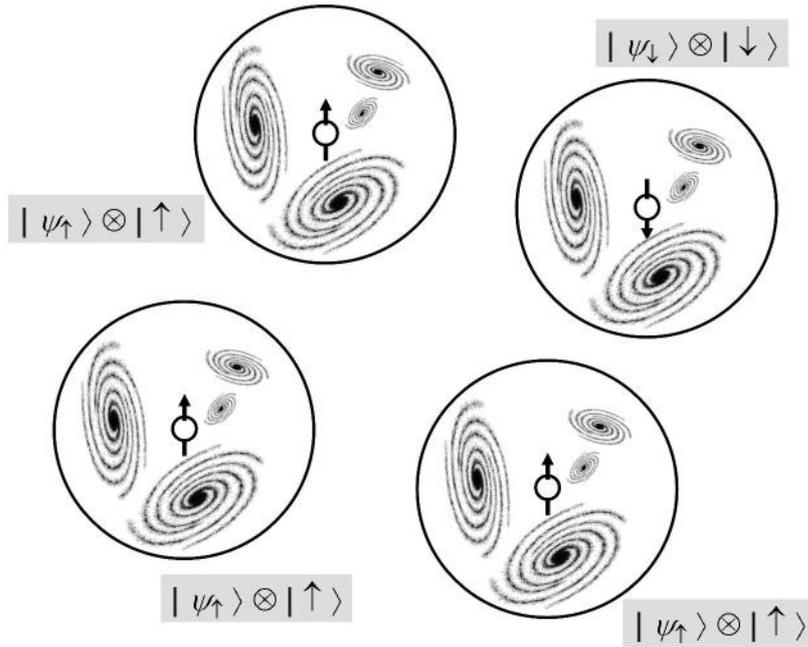

**Figure 2:** Quantum uncertainty arises concerning the outcome of the experiment – spin-up or spin-down? – because the observer in each region does not know which region she is in.

How do we know that the Aguirre-Tegmark structure is coherent? We know this because it supports Born's rule. Whether or not the bubble universe is infinite, its structure must be organized so that the frequency distribution of quantum events is in conformance with the Born rule, as in Figure 2. In order to achieve this distribution, the entire structure must evolve in accordance with the Schrödinger equation (as well as with other equations including, particularly, those of general relativity).

As Aguirre and Tegmark acknowledge, individual quantum events that are distributed throughout the bubble universe are not isolated from the environments in which they occur: they are replicated along with, for instance, identical versions of the environment in which the quantum event is manifest (the Stern-Gerlach apparatus, for instance). So we are thinking of volumes, or regions, rather than individual quantum events being replicated.

In order to validate the Aguirre-Tegmark model, we start by looking at the characteristics of the replicated regions. While the properties of each region's constituent particles (spin, position, momentum, etc.) may appear to be randomly distributed, they must nevertheless result from the evolution of the Schrödinger equation, which may be taken to determine the complete state within such a region. So, as well as being part of a coherent, structured bubble universe, each region inside the bubble universe is itself a coherent mathematical structure. In the current context, this is equivalent to saying that each region can be considered to be in a single quantum state, which, just after the outcome of a quantum event, we can consider as an eigenstate which is itself the product of all of the eigenstates in its history.

To see what this means, look at Figure 3, adapted from reference [18]. The tree in this figure schematically illustrates the history of the unitary evolution of the Schrödinger equation to generate a myriad of different universes, many of which contain replicated versions of ourselves. This, of course, is along the lines of MWI. The figure shows eight different configurations of "worlds", or universes, labelled (1)-(8), that emerge from applying the Schrödinger equation to three quantum events, A, B and C, with probability amplitudes $a$, $b$ and $c$ respectively. Event A has two possible outcomes, eigenstates $|A_1\rangle$ and $|A_2\rangle$ with probability amplitudes $a_1$ and $a_2$. Event C is contingent on $A_2$ occurring – if the outcome of event A is $A_1$ then event C does not happen. Event C has three possible outcomes, eigenstates $|C_1\rangle$, $|C_2\rangle$ and $|C_3\rangle$. Event B occurs independently of A and C and so it appears as a component in every one of the eight configurations of universes.

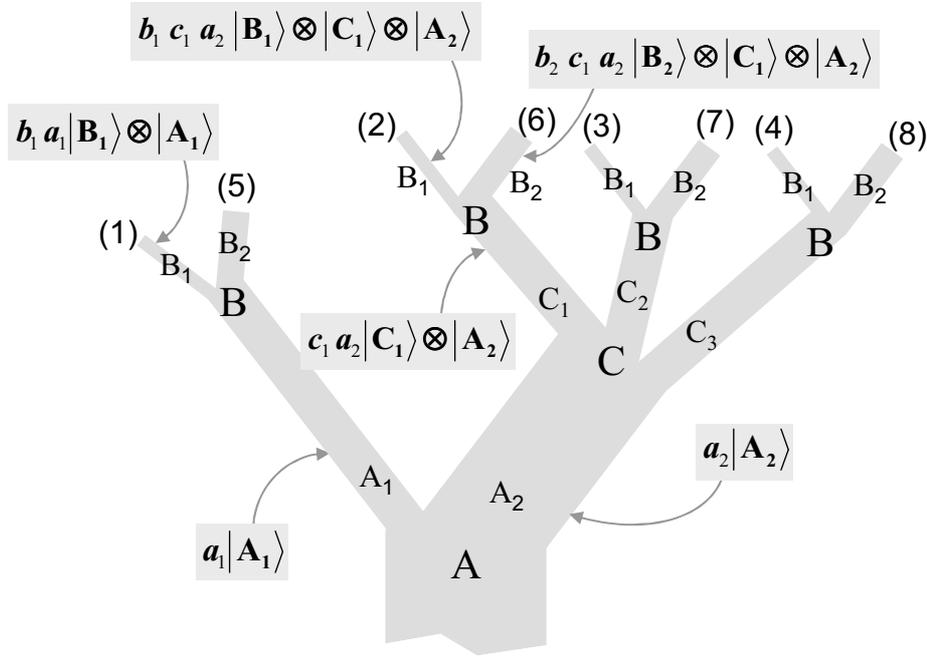

**Figure 3:** Three quantum events, A, B and C, have outcomes $|A_1\rangle$, $|A_2\rangle$, $|B_1\rangle$, $|B_2\rangle$, $|C_1\rangle$, $|C_2\rangle$ and $|C_3\rangle$ respectively, with corresponding probability amplitudes $a$, $b$ and $c$. Each branch, at every level, is an eigenstate.

So, in the Aguirre-Tegmark model for the three quantum events A, B and C, the bubble universe has to be populated by regions that are distributed in eight different configurations corresponding to those labelled (1)-(8) in Figure 3 with frequencies in proportion to the absolute squares of the probability amplitudes for these eight configurations. Hence, for instance, the frequency with which configuration (1) (*i.e.*, $b_1 a_1 |B_1\rangle \otimes |A_1\rangle$) occurs relative to configuration (6) (*i.e.*, $b_2 c_1 a_2 |B_2\rangle \otimes |C_1\rangle \otimes |A_2\rangle$) is $(b_1 a_1)^2 / (b_2 c_1 a_2)^2$.

It is important to keep in mind that each of the universes (1)-(8) is an eigenstate. Indeed, the quantum state of every branch of the tree at every level is an eigenstate, and the outcomes for each quantum event (for instance, $|A_1\rangle$ and $|A_2\rangle$) form an orthonormal basis for that event.

So each spatial region in the Aguirre-Tegmark model corresponds to a particular quantum eigenstate that emerges from the history of the outcomes of all of the events leading to that specific configuration. We can determine the spatial extent of this eigenstate within the bubble universe using a thought experiment. We begin with the premise that each region must be at least as large as a Hubble sphere because of the following argument.

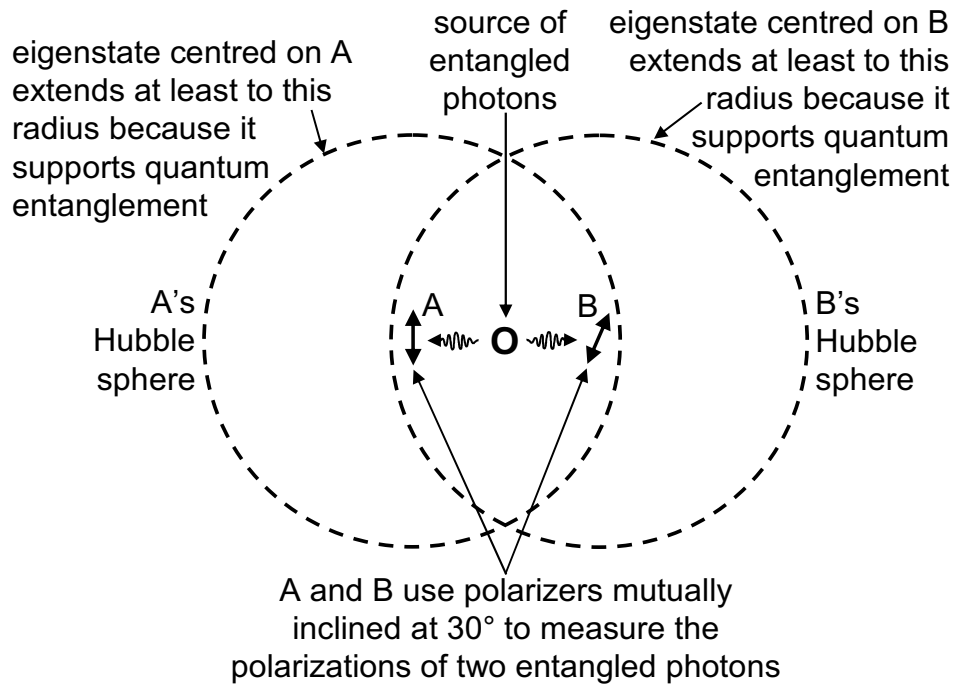

**Figure 4:** Since A can conduct an entanglement experiment with an experimenter B in any direction up to the boundary of her Hubble sphere, she concludes that the quantum state extends at least that far. B comes to the same conclusion about his own Hubble volume.

### 3. Physical size of the eigenstate

The boundary of a Hubble sphere with an observer at the centre is defined to be the distance at which expanding space recedes from the observer at the speed of light. Imagine two observers, A and B, each of whom is just within the other's Hubble sphere (see Figure 4). Suppose that they each measure the polarization of one of a pair of entangled photons, which were emitted from a source half-way between them in opposite directions at an early epoch in the universe. A and B measure the polarizations of their respective photons with polarizers that are mutually inclined at an angle of 30°, and each expects to find a polarization either vertically or horizontally inclined to their particular polarizer.

If they repeat this experiment many times with a series of pairs of entangled photons, then A and B will each expect to find a 50:50 ratio of vertical-to-horizontal polarizations. However, since each pair of photons is entangled and their polarizers are mutually at 30°, A and B also anticipate that ¾ of their results will agree (both polarizations vertical or both horizontal) with ¼ disagreeing (one polarization vertical when the other is horizontal).

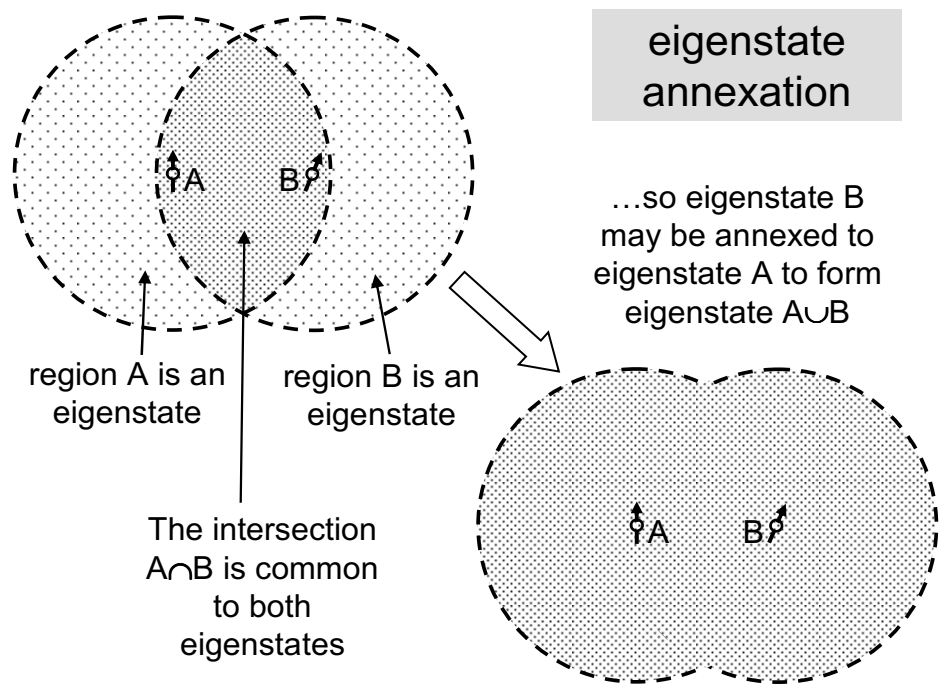

**Figure 5:** The top drawing is the same as in Figure 4. Since the eigenstate of A overlaps – is shared with – the eigenstate of B, the complete space must be in a common eigenstate. It is convenient to have a name for this process of using a common space to show that two overlapping states are in a single eigenstate and so we use the term *eigenstate annexation*.

However, A and B will not be able to verify this entanglement correlation until they have been able to exchange signals. If A and B had been separated by a distance greater than their respective Hubble radii, such exchange of signals might still be possible, depending upon the model of expanding universe[‡] (in that case, A's Hubble radius would have to increase eventually to enclose B, and vice versa). Since we suppose, though, that A and B are already (just) within each other's Hubble radius, no controversy arises in saying that they will, in due course, be able to exchange signals.

The exchange of signals confirms the entanglement, and so we may conclude that the physical extent of the quantum state $|\psi\rangle$ is (at least) as wide as a Hubble radius.

---

[‡] The radius of the Hubble sphere increases in decelerating universes, and, in accelerating universes, the radius also tends to increase. If this radius increases faster than the net recession velocity of photons immediately outside of the Hubble sphere that were emitted in the direction towards the centre of the Hubble sphere, then these photons will eventually reach their target [19].

Since observer A can conduct this entanglement experiment in any direction, she concludes that she is at the centre of a spherical region described by one of the eigenstates of the entanglement that extends as far as the boundary of her Hubble sphere (see Figure 4). Equally, of course, observer B will find that he is at the centre of a region extending, in turn, to the boundary of his own Hubble sphere.

However, since A and B share a large volume of space in common, the Hubble spheres centred around both A and B must both belong to the same quantum eigenstate, which implies that the eigenstate extends more widely than a Hubble radius (see Figure 5).

In summary, A finds herself at the centre of an eigenstate that extends to the boundary of her Hubble sphere, as does B. Since A and B share a region of space in common, A has effectively annexed the space containing B's eigenstate into a single, larger eigenstate, as we see in Figure 5. It will be convenient later to refer to this type of entanglement thought experiment as *eigenstate annexation*.

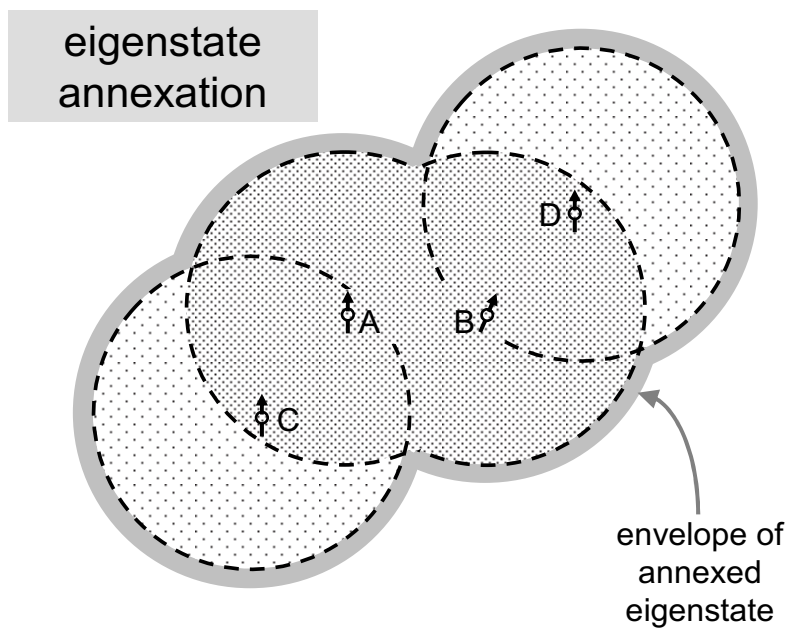

**Figure 6:** Eigenstate annexation can be extended so far that some parts of the eigenstate (*e.g.* C and D) are receding from each other at superluminal velocities because of the expansion of space.

We can extend the eigenstate annexation. Figure 6 shows a new observer, C, just within the boundary of A's Hubble sphere. A and C can conduct further entanglement experiments, which means that C can regard her Hubble sphere as part of the same eigenstate as A's eigenstate. Similarly, D, who is at the edge of B's Hubble sphere finds the same. So the whole volume enclosed by the grey outline in Figure 6 is in the same eigenstate. Notice that parts of the eigenstate, including, for instance, C and D, are no longer within a Hubble radius of each other – indeed, they are more than a Hubble diameter apart – and so their mutual recession velocity exceeds that of light[§].

It may seem counterintuitive that parts of the bubble universe that are receding from each other at superluminal velocities, and which may well be forever beyond contact with each other, are nevertheless part of the same quantum state. Another thought-experiment may help to make the idea more plausible. In Figure 7, A and B, along with their polarization detectors, are now separated by a larger distance than before: they are diametrically opposite each other, just inside the boundary of a Hubble sphere centred at observer O. Observer O is half-way between A and B, so that the distance between A and B is now nearly two Hubble radii rather than just one radius as in the previous experiments.

Observer O sends out pairs of entangled photons towards A and B who record the outcomes of their polarization measurements and return the readings to O. Since both A and B lie within O's Hubble radius, O will eventually receive their signals, which confirm the expected entanglement correlations.

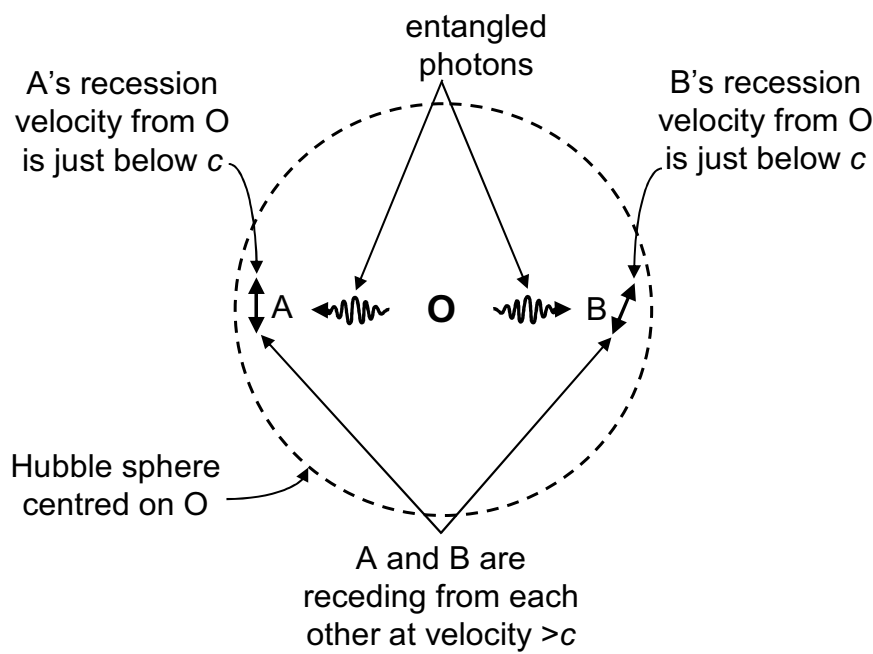

**Figure 7:** A and B are receding from each other at superluminal velocities[§] because they are separated by more than a Hubble radius. Nevertheless, records of the entanglement measurements received back at O confirm that they share the same quantum eigenstate.

A and B, being separated by nearly two Hubble radii, are receding from each other at greater than the speed of light, owing to the expansion of space. They may well never be in contact with each other, and yet the outcomes of their polarization measurements are correlated according to the usual cosine-squared rule for entangled photons, as O can (eventually) verify.

---

[§] Of course, this does not conflict with special relativity because the recession velocities are entirely the result of the general relativistic expansion of space.

While this thought experiment may make it easier to accept that a quantum state can extend across regions of space that are receding from each other at superluminal velocities, it still begs an explanation.

The photons in each entangled pair have no polarization until they are measured – there are no hidden variables, as a Bell experiment would confirm. The polarization measurements themselves are as causally separated as could ever be imagined. It may be tempting to try to explain quantum entanglement beyond a Hubble radius by looking back to the instant before inflation stretched the constituent parts of the embryonic bubble universe out of mutual causal contact. But that would be to regard the pre-inflation primordial soup as the incubator for the quantum state of the whole bubble universe. Of course, that is the wrong way round: it is the bubble universe itself that evolves according to the quantum state – the coherent mathematical structure – that describes it. In the final analysis, entanglement beyond a Hubble radius may be perhaps the supreme example of the suggestion that, in Tegmark's words [20], "our reality isn't just *described* by mathematics – it *is* mathematics".

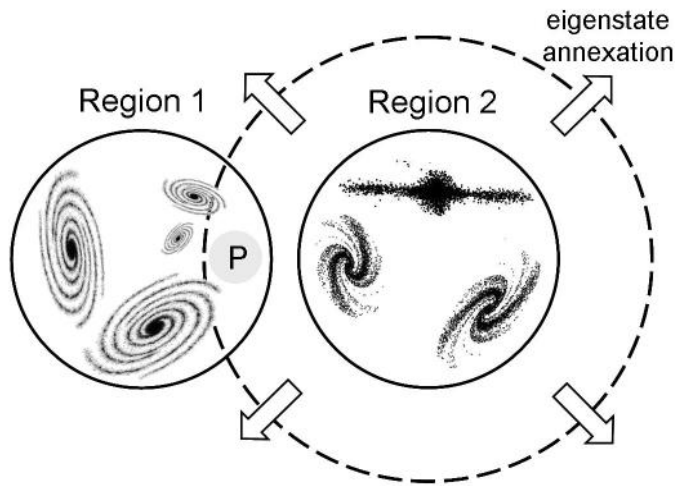

**Figure 8:** Region 1 and Region 2 may be thought of as universes (1) and (6) of Figure 3. The boundary of Region 2 is extended to perform eigenstate annexation of Region 1.

From the point of view of any observer at the edge of an eigenstate, such as C or D in Figure 6, the large-scale properties of the bubble universe appear to be the same in all directions, in accordance with Einstein's Cosmological Principle. So, however large the space containing the eigenstate appears to be, it can always be extended in any direction by annexation, and this is a process that can be repeated indefinitely. So the eigenstate extends not only beyond the Hubble radii of A and B: it extends across the whole bubble universe. In other words, the coherent mathematical structure of the whole bubble universe may be considered to be one eigenstate. (We shall see later that the complete bubble universe is effectively what we think of as "the universe", and that all of the other possible eigenstates are represented by parallel universes.)

## 4. The bubble universe is in a single eigenstate

It is difficult to reconcile this conclusion with the Aguirre-Tegmark model, represented in Figure 2, in which the 3-space of the bubble universe should contain the very many different regions required to support quantum uncertainty leading to different outcomes. To illustrate this point, look at Figure 8, which shows two different regions in the bubble universe, Region 1 and Region 2, as required by the Aguirre-Tegmark model – we might think of them, for instance, as universes (1) and (6) from Figure 3. Their different histories are indicated schematically by two different arrangements of galaxies, each bounded by a solid-line circle drawn around the two regions.

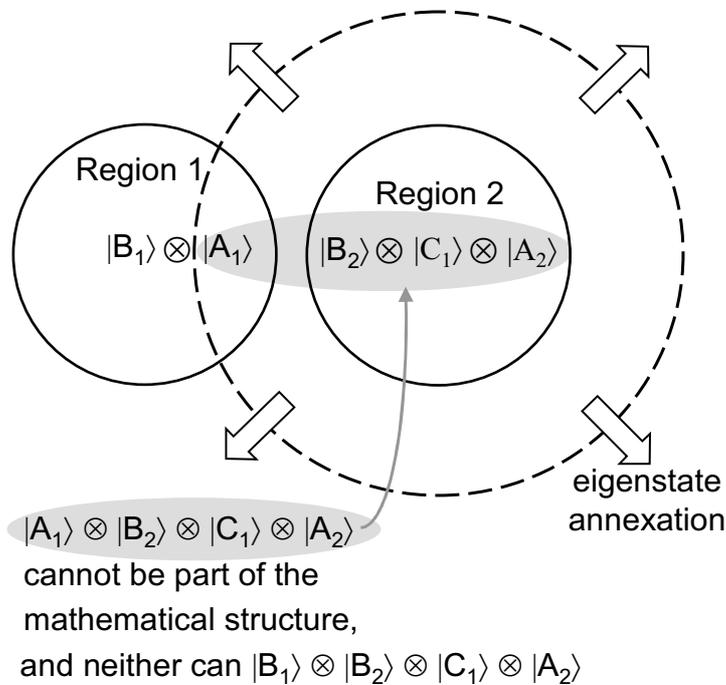

**Figure 9:** The eigenstate of Region 1 cannot be annexed into the eigenstate of Region 2 because that would involve annexing mutually orthogonal outcomes.

By the process of eigenstate annexation, Region 2 may be extended to overlap Region 1 as shown by the block arrows expanding the dotted circle. In particular, this annexation now includes point P contained within Region 1. Since point P belongs to Region 1, it is part of the quantum eigenstate, or coherent mathematical structure, of Region 1. However, point P is also encompassed by eigenstate annexation from Region 2 and so it is now also part of the quantum eigenstate, or coherent mathematical structure, of Region 2. However, because of their different histories, the mathematical structures of the two regions are in general different; indeed, they are orthogonal eigenstates. So, if event P is a logical consequence of the mathematical structure of Region 1 (and, therefore, part of it), it will not in general be a logical consequence of the mathematical structure of Region 2, and vice versa. Essentially, of course, what we are saying is that the bubble universe cannot be in two different eigenstates.

To take a specific example, consider Figure 9. The quantum state of Region 2, within the solid-line circle on the right, is eigenstate $|B_2\rangle \otimes |C_1\rangle \otimes |A_2\rangle$, which corresponds to universe (6) in Figure 3. That of Region 1, on the left, is $|B_1\rangle \otimes |A_1\rangle$, corresponding to universe (1) in Figure 3. If Region 2 is extended in an eigenstate annexation, it overlaps Region 1 as it did in Figure 8. If outcome $|A_1\rangle$ now falls within the compass of Region 2, then all of the outcomes $|A_1\rangle$, $|B_2\rangle$, $|C_1\rangle$ and $|A_2\rangle$ must be part of the mathematical structure of Region 2. However, the two possible outcomes of event A, namely $|A_1\rangle$ and $|A_2\rangle$, cannot both be present in Region 2 because they are mutually orthogonal results of quantum event A. The same conclusion applies if it is, instead, $|B_1\rangle$ that falls within the compass of Region 2.

The above discussion illustrates the general idea that, if the 3-space bubble universe is a coherent mathematical structure, then it can be in only a single eigenstate. This means that it cannot contain different regions in different eigenstates, each representing a different universe. It is difficult to see how to reconcile this with the suggestion that the Level I Multiverse is the same as the Level III Multiverse.

**5.    A multiverse of parallel, bubble, block universes in Hilbert space**

And yet, how can this be? After all, if we accept (1) that quantum uncertainty arises through a multiverse of universes and (2) that the multiverse is a coherent mathematical structure, then each of the many different universes must be part of that encompassing mathematical structure. Surely this contradicts our conclusion that the bubble universe is in a single eigenstate and so cannot accommodate different regions that are themselves in different eigenstates?

The resolution of the apparent contradiction is that, while the different universes must indeed all belong to the same mathematical structure (because they are mathematically generated by that structure), nevertheless, each individual universe must be isolated from the others so that a clash of eigenstates does not arise. So, rather than regarding the bubble universe as accommodating many different universes as in the Aguirre-Tegmark model, we advocate considering the multiverse as a set of orthogonal eigenstates, each one of which is a separate bubble universe in the form of a block universe. It is important to appreciate that these orthogonal bubble universes are not extra bubble universes generated by inflation: they are parallel universes in Hilbert space.

These universes may be represented (see Figure 10, taken from reference [18]) by separate, parallel filaments running from the trunk to the top-most twigs of the tree, with the number of universes in every branch being proportional to the branch "thickness", that is, to the absolute square of the probability amplitude for the branch. Each parallel universe must be a block universe for the reasons discussed in reference [8]. Three such universes are highlighted in Figure 10: we assume here that the thickness of twig (6) is twice that of twig (2).

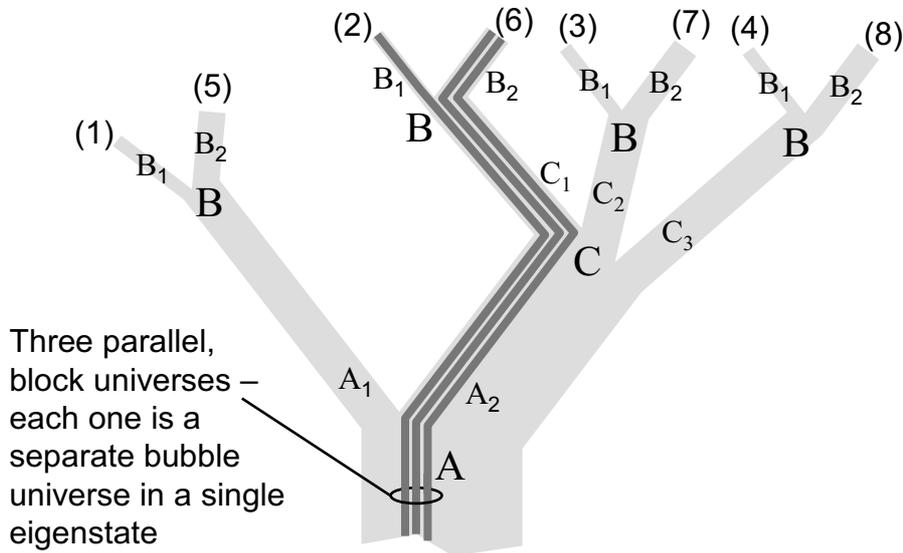

**Figure 10:** Each branch of the tree contains a number of separate, parallel block universes which is proportional to the absolute square of the probability amplitude for the branch. Each block universe is a bubble universe in a single eigenstate.

In reference [8], this multiverse of parallel block universes, along with a multitude of other multiverses, is regarded as part of a much larger mathematical structure called the *Plexus*. Apart from presenting the multiverse as a set of separate, orthogonal eigenstates, another difference between the Plexus hypothesis and the Aguirre-Tegmark model is that each block universe in the Plexus model is regarded as finite in both space and in time. (Essentially, the argument [18] is that: (1) the number of parallel universes in the multiverse is finite because probabilities cannot be derived from ratios of infinite numbers of universes; (2) the number of parallel universes in the multiverse is directly related to the total number of quantum events in the multiverse as shown in reference [18]; (3) since the number of parallel universes is finite, then so must be the total number of quantum events in any parallel universe; (4) an infinite universe will contain an infinite number of quantum events; (5) therefore, no universe in the multiverse can be infinite.)

The simplest form for a spatially finite parallel universe is the three-sphere (that is, an $S^3$ geometry), which would mean that such a universe – our own, for example – would have a positive curvature. However, data from the Planck 2015 project [21] suggest a cosmological curvature parameter of zero to within 0.5 percent. This could, of course, simply mean that the bubble universe is so much larger than the observable universe that it appears essentially flat locally but is nevertheless positively curved. In any case, space can be flat and still be finite, as in the three-torus model. Some topologies (such as Friedmann-Robertson-Walker models [22]) allow the universe to be finite and still have a zero or even negative curvature [23]. (The proposal that our spatially finite universe is finite also in time will be discussed in a subsequent paper.)

Does the proposal that a bubble universe is a block universe in a single eigenstate rule out the argument of Aguirre and Tegmark [1] that space contains very many similar or identical copies of our own Hubble volume?

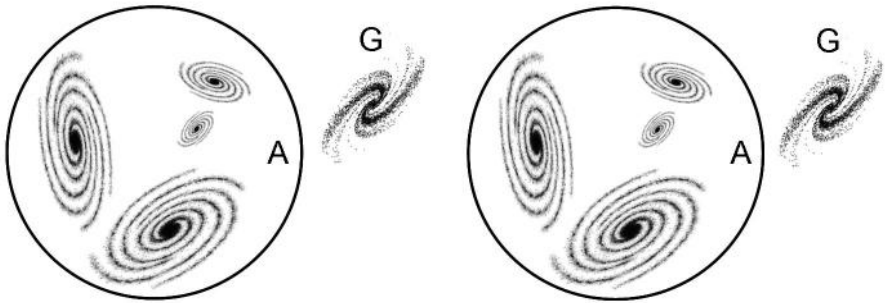

**Figure 11:** If a Hubble volume is to be replicated exactly, then so must the surrounding environment, including galaxy G, because it can be seen by observer A from the edge of the Hubble volume.

As we have seen, we have to exclude Hubble volumes in eigenstates orthogonal to that of the bubble universe, which rules out non-identical copies of our Hubble volume in our bubble universe. The only type of region that can be replicated is a Hubble volume that is completely identical to our own.

To see how the bubble universe might accommodate identical copies of our Hubble volume, look at Figure 11, in which the Hubble volume on the left is to be replicated. There is an observer, A, at the edge of the Hubble volume who can see a galaxy, G, beyond the Hubble volume. If we are to replicate the Hubble volume, we have to replicate the observer, along with her observations, and so we have to replicate the environment that she sees. This is why an identical galaxy G has been drawn to the right of the replicated Hubble volume, along with the replicated observer A.

In other words, it is not just the Hubble volume that is replicated, but the space between the spherical volumes also. In Figure 12, the attempt to represent replication in all three dimensions looks like a carpet with a repeating pattern of period of $\Delta$.

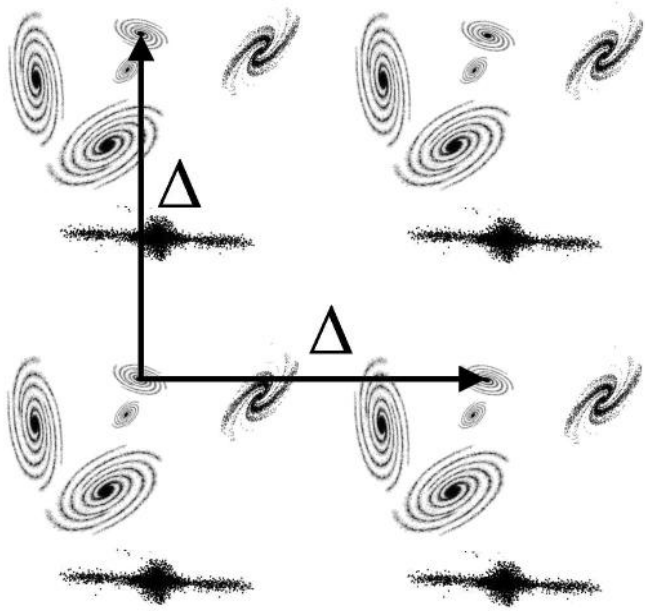

**Figure 12:** When the Hubble volumes are replicated as in Figure 11, so too is the surrounding environment. This leads to a patterned-carpet structure repeated in every spatial direction with a period Δ.

Of course, such a repeating pattern is an intrinsic characteristic of the three-sphere or the three-torus geometry that we already proposed for our bubble universe earlier in this section, and so the notion of the Hubble volume being replicated in our universe really adds nothing to the picture, nor contributes to quantum uncertainty.

## 6. Nature of the mathematical structure

There is, perhaps, a difference between the Aguirre-Tegmark model and the Plexus hypothesis that is more significant than whether parallel universes are in the same Euclidean three-space (in the Aguirre-Tegmark model) or not (in the Plexus hypothesis). It is that the mathematical structure of the Plexus is assumed to be sufficiently complex that it is incomplete, whereas Tegmark [5] avoids such Gödelian self-referential knots by allowing only a basic system of computable functions in his Computable Universe Hypothesis.

It is difficult to see how the Plexus can accommodate Schrödinger-like axioms based upon an arithmetic so simple that it escapes Gödel's net. (As discussed in references [8] and [18], in a finite multiverse, these axioms need to be compatible with a discrete model of physics such as those proposed by 't Hooft [24] and Zahedi [25]). Although a system such as Presburger arithmetic, for example, is simple enough to be complete, it cannot support the Schrödinger relations because it has no axioms for multiplication. Repeated addition cannot substitute for multiplication in the general case, and the gap is not filled, for instance, by including an axiom to define multiplication by zero, because then you move into Peano-like axioms which then fall within Gödel's compass. So, at the level of our multiverse, the Plexus appears to be incomplete.

Furthermore, it is not clear that Tegmark's appeal to Gödel's second incompleteness theorem gives us reason to doubt the consistency of the mathematical structure that contains our multiverse of universes. A system that is complex enough to be incomplete cannot prove that it is, itself, consistent, and Tegmark expressed concern that, in an inconsistent system, "mathematics as we know it would collapse like a house of cards". However, while a complex system cannot *prove* its own consistency, it is nevertheless perfectly possible for it to *be* consistent, and, as yet, we have no physical evidence that the mathematical structure, and hence our multiverse, is anything other than consistent.

The incompleteness in the Plexus is in fact resolved at the topmost (transfinite) level. As Gödel suggested in footnote 48a of his paper [26], incompleteness at any level can be accommodated by adding appropriate axioms to form a higher-level system at the expense of creating incompleteness at that higher level, and this "formation of ever higher types can be continued into the transfinite". In effect, what this means is that inhabitants of Tegmark's universe ultimately have the tools to explain it completely whereas those of universes in the Plexus may speculate but can never be sure.

Much more important, though, than the differences between Tegmark's structure and the Plexus is the very fact that they are, indeed, mathematical structures. The point about the "unquestionably real" comment of Garriga and Vilenkin [14] is that they were questioning the reality of the parallel worlds in the Many Worlds Interpretation. However, as shown in [8], even a basic mathematical structure consisting of only three lines is powerful enough for a Universal Turing Machine to emerge within it, capable of detecting, modelling and so being aware of its environment including, in principle, its own existence. To such an emergent substructure, its world is most certainly "real", and this applies to any part of the overarching mathematical structure that is capable of supporting such self-awareness. At the deepest level, it is the mathematical structure that is at the very heart of reality.